\documentclass[12pt]{article}

\usepackage{bbm}
\usepackage{amssymb,amsmath}

\allowdisplaybreaks

\newcommand{\be}{\begin{equation}}
\newcommand{\ee}{\end{equation}}
\newcommand{\ba}{\begin{eqnarray}}
\newcommand{\ea}{\end{eqnarray}}
\newcommand{\no}{\nonumber\\}
\newcommand{\zz}{\mathbbm{Z}_2}
\newcommand{\mnu}{\mathcal{M}_\nu}
\newcommand{\hm}{\hat M}

\begin{document}

\title{
\normalsize \hfill UWThPh-2009-13 \\
\normalsize \hfill CFTP/09-034 \\*[8mm]
\LARGE Trimaximal lepton mixing with a trivial Dirac phase}

\author{
W.~Grimus,$^{(1)}$\thanks{E-mail: walter.grimus@univie.ac.at}
\
L.~Lavoura$\,^{(2)}$\thanks{E-mail: balio@cftp.ist.utl.pt}
\
and A.~Singraber$\,^{(1)}$\thanks{E-mail: a0403346@unet.univie.ac.at}
\\*[3mm]
$^{(1)} \! $
\small University of Vienna, Faculty of Physics \\
\small Boltzmanngasse 5, A--1090 Vienna, Austria
\\*[2mm]
$^{(2)} \! $
\small Technical University of Lisbon, Inst.\ Sup.\ T\'ecnico, CFTP \\
\small 1049-001 Lisbon, Portugal
}

\date{12 January 2010}

\maketitle

\begin{abstract}
We present a model which employs the seesaw mechanism
with five right-handed neutrinos,
leading to trimaximal and $CP$-conserving lepton mixing. 
Tri-bimaximal mixing is a natural limiting case of our model
which occurs when one particular vacuum expectation value is real 
and preserves the $\mu$--$\tau$ interchange symmetry of the Lagrangian.
Our model allows for leptogenesis
even in the case of exact tri-bimaximal mixing.
\end{abstract}

\newpage

\section{Introduction}

It is well known that the Standard Model (SM)
of the electroweak interactions is incomplete,
because (among other reasons) it offers
neither an explanation for the exceptionally small neutrino masses
nor for the generation of a baryon asymmetry of the Universe (BAU)
of the observed size.
One possible way out of these problems
consists in the introduction of gauge-invariant right-handed neutrinos.
Their Majorana mass terms
are not proportional to the vacuum expectation values (VEVs)
which break the gauge symmetry and may,
therefore,
be of a very high mass scale.
This leads to a seesaw mechanism~\cite{seesaw}
generating masses for the standard left-handed
neutrinos---masses which are,
as a consequence of the high mass scale, 
strongly suppressed.
The extra-heavy right-handed neutrinos also allow
for the generation of the BAU
through the mechanism of leptogenesis~\cite{leptogenesis1,leptogenesis2},
which is based on
$CP$-violating asymmetries
in their Yukawa-interaction-mediated decays.

On the experimental side~\cite{exp},
it is by now established that
there is neutrino mixing in the weak charged current.
This should,
according to theory,
be parameterized
by a $3 \times 3$ unitary mixing matrix $U_\mathrm{PMNS}$.
Although the data are not sufficiently precise yet,
it appears that this matrix is close
to the tri-bimaximal mixing (TBM) form~\cite{HPS}:
\be
\label{HPS}
U_\mathrm{PMNS} \approx U_\mathrm{TBM} \equiv
\left( \begin{array}{ccc}
2 \left/ \sqrt{6} \right. &
1 \left/ \sqrt{3} \right. &
0 \\ 
- 1 \left/ \sqrt{6} \right. &
1 \left/ \sqrt{3} \right. &
- 1 \left/ \sqrt{2} \right. \\ 
- 1 \left/ \sqrt{6} \right. &
1 \left/ \sqrt{3} \right. &
1 \left/ \sqrt{2} \right.
\end{array} \right),
\ee
where we have omitted
the possible presence of non-zero Majorana phases
multiplying $U_\mathrm{TBM}$ on the right.
A milder assumption is that $U_\mathrm{PMNS}$ only displays
trimaximal mixing (TM),
which is defined as the second column of $U_\mathrm{PMNS}$
being $\left( 1, 1, 1 \right)^T \! \left/ \sqrt{3} \right.$.

There exist in the literature models based on the seesaw mechanism
which predict TBM~\cite{TBM1,prescient,tribi} or TM~\cite{tri,tri1}.
Unfortunately,
models predicting TBM~\cite{TBM1}
usually do not allow for leptogenesis,
as was recently pointed out by Jenkins and Manohar~\cite{manohar}.
For recent studies of the relationship between TBM
(or rather deviations from TBM)
and leptogenesis see~\cite{hagedorn}.

In this paper we propose a generalization
of a previous model of two of us~\cite{tribi} which predicted TBM.
The generalization leads to TM
with the additional prediction of a real (except for Majorana phases)
$U_\mathrm{PMNS}$,
thus allowing for greater predictivity than just TM.
We moreover demonstrate that our model,
both in its TM and in its TBM versions,
allows for leptogenesis.

\section{The model}

The model discussed here was introduced in~\cite{tribi}
and is an extension of the SM based on the gauge group $SU(2) \times U(1)$.
Using the index $\alpha = e,\mu,\tau$,
the fermion sector consists of $SU(2)$ doublets $D_{\alpha L}$,
$SU(2)$ singlets $\alpha_R$ with electric charge $-1$
and $SU(2)$ singlets $\nu_{\alpha R}$ with electric charge $0$. 
There are two additional
right-handed neutrino singlets $\nu_{jR}$ ($j = 1, 2$),
so that the total number of right-handed neutrinos is five.
The scalar sector consists of four Higgs doublets
$\phi_0$ and $\phi_\alpha$,
all of them with weak hypercharge 1/2,
and a complex gauge singlet $\chi$.

The family symmetries of our model are the following:
\begin{itemize}
\item Three $U(1)$ symmetries
associated with the family lepton numbers $L_\alpha$.
All scalar fields have $L_\alpha = 0$ for all $\alpha = e, \mu, \tau$.
The fermion multiplets $D_{\beta L}$,
$\beta_R$ and $\nu_{\beta R}$
have lepton number $L_\alpha = 1$ if $\beta = \alpha$
and $L_\alpha = 0$ otherwise.
\item Three $\mathbbm{Z}_2$ symmetries~\cite{prescient,nasri}
\be
\label{Z2}
\mathbbm{Z}_2^{(\alpha)}: \quad \alpha_R \to - \alpha_R, \
\phi_\alpha \to - \phi_\alpha,
\ee
for $\alpha = e, \mu, \tau$.
\item The permutation symmetry $S_3$ of the indices $e,\mu,\tau$.
With respect to this $S_3$ the gauge multiplets are arranged in triplets,
doublets and one singlet as
\be
\label{multiplets}
\left( \begin{array}{c} 
D_{e L} \\ D_{\mu L} \\ D_{\tau L} \end{array} \right), \quad
\left( \begin{array}{c} 
e_R \\ \mu_R \\ \tau_R \end{array} \right), \quad
\left( \begin{array}{c} 
\nu_{e R} \\ \nu_{\mu R} \\ \nu_{\tau R} \end{array} \right), \quad
\left( \begin{array}{c} 
\phi_e \\ \phi_\mu \\ \phi_\tau \end{array} \right), \quad
\left( \begin{array}{c} 
\nu_{1 R} \\ \nu_{2 R} \end{array} \right), \quad
\left( \begin{array}{c} 
\chi \\ \chi^* \end{array} \right), \quad
\phi_0,
\ee
respectively. 
We view $S_3$ as being generated
by the $\mu$--$\tau$ interchange $I_{\mu\tau}$
and the cyclic permutation $C_{e\mu\tau}$,
which are represented by
\be
I_{\mu\tau} \to 
\left( \begin{array}{ccc}
1 & 0 & 0 \\ 0 & 0 & 1 \\ 0 & 1 & 0 \end{array} \right), 
\quad
C_{e\mu\tau} \to \left( \begin{array}{ccc}
0 & 1 & 0 \\ 0 & 0 & 1 \\ 1 & 0 & 0 \end{array} \right)
\ee
in the case of triplets and by
\be
I_{\mu\tau} \to 
\left( \begin{array}{ccc}
0 & 1 \\ 1 & 0 \end{array} \right), 
\quad
C_{e\mu\tau} \to 
\left( \begin{array}{cc}
\omega & 0 \\ 0 & \omega^2 \end{array} \right)
\ee
for the doublets,
where $\omega = \exp (2i\pi/3)$.
\end{itemize}
The symmetries defined here generate a group which has the structure
of a semidirect product, with $S_3$ acting upon the three $U(1)$ and
the three $\zz$ symmetries---for details see~\cite{tribi}. 
Under this group the multiplets of~(\ref{multiplets}) are irreducible.

\begin{table}[t]
\be\nonumber
\begin{array}{c|cc}
\mbox{symmetry} & \mbox{dimension} & \mbox{Lagrangian} \\ \hline 
U(1)_{L_\alpha} & 3 & \mathcal{L}_\mathrm{Majorana} \\
\mathbbm{Z}_2^{(\alpha)} & 2 & V \\
C_{e\mu\tau} & 2,1 & V
\end{array}
\ee
\caption{Soft breaking of the family symmetries of our model.
In the first column the symmetry is indicated,
the second column gives the dimension of the terms
responsible for the soft breaking
and the third column specifies the part of the
Lagrangian where the soft breaking occurs.
$\mathcal{L}_\mathrm{Majorana}$
refers to the mass terms of the right-handed neutrino singlets;
$V$ refers to the scalar potential.
$I_{\mu\tau}$ is not softly broken, only spontaneously.
\label{soft breaking}}
\end{table}
In the model there is both soft and spontaneous symmetry breaking.
The soft breaking proceeds stepwise
as specified in table~\ref{soft breaking}.
The symmetry $I_{\mu\tau}$ is not broken softly.
The lepton numbers are softly broken at high energy,
\textit{i.e.}~at the seesaw scale~\cite{GL01}
where the right-handed neutrino singlets acquire Majorana mass terms.
The symmetries $\mathbbm{Z}_2^{(\alpha)}$ and $C_{e \mu \tau}$
are softly broken at low energy,
\textit{i.e.}~at the electroweak scale.
All the symmetries except the family lepton numbers
are spontaneously broken.

The multiplets and symmetries uniquely 
determine the Yukawa Lagrangian
\begin{subequations}
\label{yuwa}
\ba
\mathcal{L}_\mathrm{Yukawa} &=&
- y_1 \sum_{\alpha = e, \mu, \tau } \bar D_{\alpha L} \alpha_R \phi_\alpha
\label{alphar} \\ & &
- y_2 \sum_{\alpha = e, \mu, \tau}
\bar D_{\alpha L} \nu_{\alpha R}
\left( i \tau_2 \phi_0^\ast \right)
\label{nualphar} \\ & &
+ \frac{y_3}{2} \left( \chi \, \nu_{1R}^T C^{-1} \nu_{1R} +
\chi^\ast \, \nu_{2R}^T C^{-1} \nu_{2R} \right)
+ \mathrm{H.c.}
\label{chi}
\ea
\end{subequations}
Note that this Yukawa Lagrangian has a minimal number of couplings.
The charged-lepton masses 
$m_\alpha = \left| y_1 v_\alpha \right|$ ($\alpha = e,\mu,\tau$)
are different because of the different VEVs
$v_\alpha \equiv \left\langle \phi^0_\alpha \right\rangle_0$;
these different VEVs of course break spontaneously
both $I_{\mu \tau}$ and $C_{e \mu \tau}$.

Taking into account the pattern of soft symmetry breaking
outlined in table~\ref{soft breaking},
the Majorana mass terms of the right-handed neutrino singlets are given by
\begin{subequations}
\label{Maj}
\ba
\mathcal{L}_\mathrm{Majorana} &=&
\frac{M_0^\ast}{2} \sum_{\alpha = e, \mu, \tau}
\nu_{\alpha R}^T C^{-1} \nu_{\alpha R}
\label{m0} \\ & &
+ M_1^\ast \left(
\nu_{eR}^T C^{-1} \nu_{\mu R}
+ \nu_{\mu R}^T C^{-1} \nu_{\tau R}
+ \nu_{\tau R}^T C^{-1} \nu_{eR}
\right)
\label{m1} \\*[1mm] &&  
+ M_2^\ast \left[ \nu_{1R}^T C^{-1} \left(
\nu_{eR} + \omega \nu_{\mu R} + \omega^2 \nu_{\tau R} \right)
+ \nu_{2R}^T C^{-1} \left(
\nu_{eR} + \omega^2 \nu_{\mu R} + \omega \nu_{\tau R} \right)
\right] \hspace*{5mm}
\label{m2} \\*[1mm] & &
+ M_4^\ast \nu_{1R}^T C^{-1} \nu_{2R}
+ \mathrm{H.c.}
\label{LM}
\ea
\label{majorana}
\end{subequations}

\section{Neutrino masses and lepton mixing}

It is easy to derive the $5 \times 3$ neutrino Dirac mass matrix $M_D$
from~(\ref{nualphar})
and to derive
the $5 \times 5$ right-handed-neutrino Majorana mass matrix $M_R$
from~(\ref{chi}) and~(\ref{Maj}).
One obtains
\be
\label{MD}
M_D = \left( \begin{array}{ccc}
a & 0 & 0 \\ 0 & a & 0 \\ 0 & 0 & a \\
0 & 0 & 0 \\ 0 & 0 & 0
\end{array} \right),
\quad 
M_R = \left( \begin{array}{ccccc}
M_0 & M_1 & M_1 & M_2 & M_2 \\
M_1 & M_0 & M_1 & \omega^2 M_2 & \omega M_2 \\
M_1 & M_1 & M_0 & \omega M_2 & \omega^2 M_2 \\
M_2 & \omega^2 M_2 & \omega M_2 & M_N & M_4 \\
M_2 & \omega M_2 & \omega^2 M_2 & M_4 & M^\prime_N 
\end{array} \right),
\ee
where $a \equiv y_2^\ast v_0$
($v_0 \equiv \left\langle \phi_0^0 \right\rangle_0$)
and
\be\label{MN}
M_N \equiv y_3^\ast v_\chi^\ast, \quad
M^\prime_N \equiv y_3^\ast v_\chi
\quad 
\mbox{with} \ 
v_\chi \equiv \left\langle \chi \right\rangle_0.
\ee
We assume that $v_\chi$,
and hence $M_N$ and $M_N^\prime$,
are of the same (very high) scale as the bare Majorana masses $M_{0,1,2,4}$. 
We apply
the seesaw formula~\cite{seesaw}
to obtain a light-neutrino mass matrix
\be
\label{mnu}
\mnu = - M_D^T M_R^{-1} M_D = 
\left( \begin{array}{ccc}
x + y + t &
z + \omega^2 y + \omega t &
z + \omega y + \omega^2 t \\
z + \omega^2 y + \omega t &
x + \omega y + \omega^2 t &
z + y + t \\
z + \omega y + \omega^2 t &
z + y + t &
x + \omega^2 y + \omega t
\end{array} \right).
\ee
The precise formulas for $x$ and $z$ are found in~\cite{tribi};
here it suffices
to know that $x$ and $z$ are independent of $y$ and $t$.
On the other hand~\cite{tribi},
\begin{subequations}
\label{yt}
\ba
y &=& - a^2\, \frac{\left( M_0 + 2 M_1 \right) M_2^2}{\det{M_R}}\, M_N^\prime,
\label{y}
\\
t &=& - a^2\, \frac{\left( M_0 + 2 M_1 \right) M_2^2}{\det{M_R}}\, M_N.
\label{t}
\ea
\end{subequations}
From~(\ref{MN}) and~(\ref{yt})
one deduces that $y / t = e^{2 i \xi}$,
where $\xi \equiv \arg v_\chi$.
We therefore use the parameterization
\be
y = e^{i \xi} u, \quad t = e^{- i \xi} u,
\ee
with a complex $u$.

Since the phase of $x$ may be removed from the $\mnu$ of equation~(\ref{mnu})
without destroying its form,
that mass matrix has six real degrees of freedom:
three moduli $\left| x \right|$,
$\left| z \right|$ and $\left| u \right|$,
and three phases 
$\arg z$,
$\arg u$ and $\xi$.

If the VEV $v_\chi$ of $\chi$ is real,
then $I_{\mu\tau}$ is not spontaneously broken at the seesaw scale
and TBM ensues~\cite{tribi}.
We generalize that situation in this paper to allow
for a general $\xi$.
As we shall next see,
this leads to TM with the added bonus of a real lepton mixing matrix
(but for possible Majorana phases).

It is easy to check that the $\mnu$ of~(\ref{mnu}) is diagonalized by
\be
U^T \mnu U = \mathrm{diag} \left( \mu_1, \mu_2, \mu_3 \right),
\ee
with
\be
U = \left( \begin{array}{ccc} 
2 c \left/ \sqrt{6} \right. & 
1 \left/ \sqrt{3} \right. & 
2 s \left/ \sqrt{6} \right.
\\
- c \left/ \sqrt{6} \right. + s \left/ \sqrt{2} \right. & 
1 \left/ \sqrt{3} \right. &
- s \left/ \sqrt{6} \right. - c \left/ \sqrt{2} \right.
\\
- c \left/ \sqrt{6} \right. - s \left/ \sqrt{2} \right. & 
1 \left/ \sqrt{3} \right. &
- s \left/ \sqrt{6} \right. + c \left/ \sqrt{2} \right.
\end{array} \right),
\label{u}
\ee
where $c \equiv \cos (\xi/2)$ and $s \equiv \sin (\xi/2)$.
It is remarkable that,
despite the occurrence of three phases in $\mnu$,
the diagonalization matrix $U$ is real,
\textit{i.e.}~there is no Dirac phase.
The eigenvalues of $\mnu$,
though,
are complex,
\textit{i.e.}~there are Majorana phases:
\begin{subequations}
\begin{eqnarray}
\mu_1 &=& x - z + 3u, \\
\mu_2 &=& x + 2z, \\
\mu_3 &=& x - z - 3u. 
\end{eqnarray}
\end{subequations}
They have the same dependence on the three complex parameters
as in the TBM case.
Since in our model the charged-lepton mass matrix
is automatically diagonal,
$U = U_\mathrm{PMNS}$ but for the Majorana phases
which result from rendering $\mu_{1,2,3}$ real and positive.

If we define
\begin{subequations}
\label{definitions}
\ba
\tan^2 \theta_\odot &\equiv& \left| \frac{U_{e2}}{U_{e1}} \right|^2,
\\
\cos{2 \theta_\mathrm{atm}} &\equiv& \frac
{\left| U_{\tau 3} \right|^2 - \left| U_{\mu 3} \right|^2}
{\left| U_{\tau 3} \right|^2 + \left| U_{\mu 3} \right|^2},
\ea
\end{subequations}
then we see that in our model
these quantities are functions of $\left| U_{e3} \right|^2$:
\begin{subequations}
\label{functions}
\ba
\label{utywq}
\tan^2 \theta_\odot &=& \frac{1}{2 - 3 \left| U_{e3} \right|^2},
\\
\label{utywt}
\cos^2{2 \theta_\mathrm{atm}} &=& \left| U_{e3} \right|^2
\frac{2 - 3 \left| U_{e3} \right|^2}
{\left( 1 - \left| U_{e3} \right|^2 \right)^2}.
\ea
\end{subequations}
As always when mixing is trimaximal,
$\tan^2 \theta_\odot$ cannot be smaller than $1/2$~\cite{tri,albright};
this is slightly disfavoured by experiment~\cite{fogli,tortola}.
Equations~(\ref{functions})
are translated into numeric form
in table~\ref{numeric},
\begin{table}
\be\nonumber
\begin{array}{c|c|c}
\left| U_{e3} \right|^2 & \tan^2{\theta_\odot}
& \cos^2{2 \theta_\mathrm{atm}}
\\ \hline
0 & 0.5 & 0 \\
0.01 & 0.5076 & 0.0201 \\
0.02 & 0.5155 & 0.0404 \\
0.03 & 0.5236 & 0.0609 \\
0.04 & 0.5319 & 0.0816 \\
0.05 & 0.5405 & 0.1025
\end{array}
\ee
\caption{
The solar and atmospheric mixing angles
as functions of $\left| U_{e3} \right|^2$
in our model.
\label{numeric}}
\end{table}
which we should compare to the experimental values~\cite{tortola}
\begin{subequations}
\label{exp}
\ba
\left| U_{e3} \right|^2 &\le& 0.056 \ \left( 3 \sigma \right),
\\
\tan^2{\theta_\odot} &=& 0.437^{+0.047}_{-0.033},
\\
\cos^2{2 \theta_\mathrm{atm}} &\le& 0.02 \left( 1 \sigma \right).
\ea
\end{subequations}
We see that,
in the context of our model,
the experimental bound on $\theta_\mathrm{atm}$
places significant constraints
on both $\left| U_{e3} \right|$ and $\theta_\odot$.

As argued before,
the mass matrix~(\ref{mnu}) has six parameters;
these correspond to the three neutrino masses,
which are free in our model,
the modulus $\left| U_{e3} \right|$ and the two Majorana phases.
If one wishes,
the Majorana phases can be enforced to be trivial
because our model is compatible with the $CP$ transformation
\be
\label{CP}
CP: \quad \left\{ \begin{array}{rcl}
D_{\alpha L} &\to& i S_{\alpha \beta} \gamma^0 C \overline{D_{\beta L}}^T, \\
\alpha_R &\to& i S_{\alpha \beta} \gamma^0 C \overline{\beta_R}^T, \\
\nu_{\alpha R} &\to& i S_{\alpha \beta} \gamma^0 C
\overline{\nu_{\beta R}}^T, \\
\nu_{jR} &\to& i \gamma^0 C \overline{\nu_{jR}}^T \\
\phi_\alpha &\to& S_{\alpha \beta} \phi_\beta^\ast \\
\phi_0 &\to& \phi_0^\ast \\
\chi &\to& \chi^\ast
\end{array} \right.
\quad \mbox{with} \quad 
S = \left( \begin{array}{ccc}
1 & 0 & 0 \\ 0 & 0 & 1 \\ 0 & 1 & 0 
\end{array} \right).
\ee
The
effect of this $CP$ symmetry is to render $y_{1,2,3}$ in~(\ref{yuwa})
and $M_{0,1,2,4}$ in~(\ref{Maj}) real.
However,
in that case we would have $M_N^\ast = M_N^\prime$
which would make leptogenesis impossible, 
as one can deduce from the computations
in the next section.
Therefore,
we will not impose this $CP$ symmetry in the following.

\section{Leptogenesis}

In order to study leptogenesis
we need the masses $\hm_k$ ($k = 1, \ldots, 5$) of the heavy neutrinos
and the diagonalization matrix $V$ of $M_R$:
\be
V^T M_R V = \mathrm{diag} \left( \hm_1, \hm_2, \hm_3, \hm_4, \hm_5 \right),
\ee
where $V$ is $5 \times 5$ unitary and the $\hm_k$ are real and positive.
The relevant matrix for leptogenesis
is then $R \equiv V^T M_D M_D^\dagger V^\ast$.
Given the form of $M_D$ in~(\ref{MD}),
we find
\be
R = \left| a \right|^2 \bar V^T \bar V^\ast,
\label{R}
\ee
where $\bar V$ is the upper $3 \times 5$ submatrix of $V$:
\be
V = \left( \begin{array}{c} \bar V \\ \hat V \end{array} \right),
\quad \bar V: \ 3 \times 5 \ \mathrm{matrix},
\quad \hat V: \ 2 \times 5 \ \mathrm{matrix}.
\ee
We make the \textit{Ansatz}
\be
V = V_u V_b, \quad
V_u = \left( \begin{array}{cc} U_\mathrm{TBM} & 0 \\ 0 & U_\mathrm{BM}
\end{array} \right), \quad
\mathrm{where} \ U_\mathrm{BM} = \frac{1}{\sqrt{2}} \left( \begin{array}{cc}
1 & -1 \\ 1 & 1 \end{array} \right).
\ee
It is clear that
\be
R = \left| a \right|^2 \bar V_b^T \bar V_b^\ast,
\ee
where
\be
V_b = \left( \begin{array}{c} \bar V_b \\ \hat V_b \end{array} \right),
\quad \bar V_b: \ 3 \times 5 \ \mathrm{matrix},
\quad \hat V_b: \ 2 \times 5 \ \mathrm{matrix}.
\ee
$V_b$ is the unitary matrix that diagonalizes
\be
V_u^T M_R V_u = \left( \begin{array}{ccccc}
M_0 - M_1 & 0 & 0 & \tilde M_2 & 0 \\
0 & M_0 + 2 M_1 & 0 & 0 & 0 \\
0 & 0 & M_0 - M_1 & 0 & - i \tilde M_2 \\
\tilde M_2 & 0 & 0 & \bar M_N + M_4 & \hat M_N \\
0 & 0 & - i \tilde M_2 & \hat M_N & \bar M_N - M_4
\end{array} \right),
\label{nghyr}
\ee
where $\tilde M_2 \equiv \sqrt{3} M_2$,
$\bar M_N \equiv \left. \left( M_N^\prime + M_N \right) \right/ 2$
and $\hat M_N \equiv \left. \left( M_N^\prime - M_N \right) \right/ 2$.
One sees that $\hat M_2 = \left| M_0 + 2 M_1 \right|$.
The matrix~(\ref{nghyr}) neatly separates into two $2 \times 2$ submatrices
in the special case $M_N^\prime = M_N$ which corresponds to TBM;
we shall from now on restrict ourselves to that special case.
From~(\ref{nghyr}) one deduces that $\bar V_b$ has the structure
\be
\bar V_b = \left( \begin{array}{ccccc}
e^{i \varphi_1} \cos{\psi_1} & 0 & 0 & e^{i \varphi_2} \sin{\psi_1} & 0 \\
0 & e^{i \theta} & 0 & 0 & 0 \\
0 & 0 & e^{i \varphi_3} \cos{\psi_2} & 0 & e^{i \varphi_4} \sin{\psi_2}
\end{array} \right),
\ee
hence
\be
R = \left| a \right|^2 \left( \begin{array}{ccccc}
\cos^2{\psi_1} & 0 & 0 & n_1 & 0 \\
0 & 1 & 0 & 0 & 0 \\
0 & 0 & \cos^2{\psi_2} & 0 & n_2 \\
n_1^\ast & 0 & 0 & \sin^2{\psi_1} & 0 \\
0 & 0 & n_2^\ast & 0 & \sin^2{\psi_2}
\end{array} \right),
\ee
where $n_1 \equiv e^{i \left( \varphi_1 - \varphi_2 \right)}
\sin{\psi_1} \cos{\psi_1}$ and $n_2 \equiv
e^{i \left( \varphi_3 - \varphi_4 \right)} \sin{\psi_2} \cos{\psi_2}$.

Let us now suppose that $\hat M_1 \ll \hat M_{2,3,4,5}$.
Then,
the $CP$-violating asymmetry relevant for leptogenesis is
\be
\epsilon_1 = \frac{1}{8 \pi \left| v_0 \right|^2 R_{11}}\,
\sum_{k=1}^5 \mathrm{Im} \left[ \left( R_{1k} \right)^2 \right].
\ee
It is clear that the only non-zero contribution to the sum occurs for $k= 4$,
which means that in this case
\emph{leptogenesis involves only two heavy neutrinos},
and
\ba
\epsilon_1 &=&
\frac{\left| a \right|^2}{8 \pi \left| v_0 \right|^2}\,
\sin^2{\psi_1} \sin{\left( 2 \varphi_1 - 2 \varphi_2 \right)}
\no &=&
\frac{\left| y_2 \right|^2}{8 \pi}\,
\sin^2{\psi_1} \sin{\left( 2 \varphi_1 - 2 \varphi_2 \right)}.
\ea
Notice that this $CP$-violating asymmetry
crucially depends on the
\emph{difference $2 \varphi_1 - 2 \varphi_2$
between the Majorana phases of the two heavy neutrinos
with masses $\hat M_1$ and $\hat M_4$}.

In our model we are able to obtain a non-zero leptogenesis
even in the case,
treated above,
of TBM,
\textit{i.e.}\ even when $M^\prime_N = M_N$.
This does not conform with the observation in~\cite{manohar}
that the models in the literature that generate exact TBM 
using a flavour symmetry do not have leptogenesis.
The crucial point is that our model has \emph{more than three}
(specifically, five)
\emph{right-handed neutrinos}
and this leads to a matrix $\bar V$ in~(\ref{R}) which 
is not a unitary $3 \times 3$ matrix---in which case
$R \propto \mathbbm{1}$ would be trivial 
and leptogenesis would not be possible---but rather
a $3 \times 5$ submatrix of a unitary $5 \times 5$ matrix.
Thus,
our model model serves as an example
of how it is possible to evade the limitation pointed out in~\cite{manohar}:
having more than three right-handed neutrinos in the seesaw mechanism
allows one to reconcile TBM with leptogenesis.

It is well known that
in the most general case
the phases responsible for $CP$ violation at low energies
and those responsible for leptogenesis
are not related~\cite{casas}.
This is the case in the present model:
the Majorana phases in the diagonalization matrix of $\mnu$,
which contribute to neutrinoless double-$\beta$ decay,
and those in the diagonalization matrix $V$ of $M_R$
are not directly related. 
This is in contrast to the $\mu$--$\tau$-symmetric model of~\cite{GL-lg}
with a $3 \times 3$ matrix $M_R$ where those phases are identical.
Nevertheless,
if we apply the $CP$ transformation~(\ref{CP}) to the present model, 
one can read off from the matrix~(\ref{nghyr}) that 
both sets of phases,
for neutrinoless double-$\beta$ decay and for leptogenesis,
become trivial.

\section{Conclusions}

In this paper we have generalized the model introduced in~\cite{tribi}
by allowing for a complex VEV $v_\chi$
of the scalar gauge singlet $\chi$.
This generalization leads to trimaximal lepton mixing,
with tri-bimaximal mixing as the limiting case
when $v_\chi$ is real.
Although
the phase $\xi$ of $v_\chi$ shows up
in the light-neutrino mass matrix $\mnu$,
it does not induce a non-trivial Dirac phase in the PMNS matrix
but rather a non-zero element $U_{e3}$.
This is a special feature of the model
due to its symmetries
and to the five right-handed neutrino singlets in the seesaw mechanism.
Another property of the model,
resulting from five
instead of three right-handed neutrinos,
is the possibility of leptogenesis even in the limit of exact TBM.
It is worth noting that,
like its precursor in~\cite{tribi},
the model has a minimal number of Yukawa couplings;
the price to pay is that
the different charged-lepton masses have to be generated by different VEVs,
which in turn needs a sufficiently rich scalar potential. 
As for the renormalization-group evolution of $\mnu$,
it is possible to achieve stability under the one-loop evolution
through a slightly different choice of symmetries,
in the same way as discussed in~\cite{tribi,tri1}. 

Since the model predicts trimaximal mixing
and thus $\sin^2 \theta_\odot \geq 1/3$,
where
$\theta_\odot$ is the solar mixing angle,
there is a slight tension
with the present fits to the neutrino-oscillation data
and the model can probably be tested in the near future,
also by taking into account the correlations,
given in~(\ref{functions}),
between the three mixing angles.

\paragraph{Acknowledgements:}
We acknowledge support from the European Union
through the network programme MRTN-CT-2006-035505.
The work of L.L.~was supported by the Portuguese
\textit{Funda\c c\~ao para a Ci\^encia e a Tecnologia}
through the project U777--Plurianual.

\newpage


\begin{thebibliography}{99}

\bibitem{seesaw}
P.~Minkowski,
\textit{$\mu \to e \gamma$ at a rate of one out of $10^9$ muon decays?},
\textit{Phys. Lett.} \textbf{67B} (1977) 421;
\\
T.~Yanagida,
\textit{Horizontal gauge symmetry and masses of neutrinos},
in \textit{Proceedings of the workshop on unified theory
and baryon number in the universe (Tsukuba, Japan, 1979)},
O.~Sawata and A.~Sugamoto eds.,
KEK report \textbf{79-18},
Tsukuba, 1979;
\\
S.L.~Glashow,
\textit{The future of elementary particle physics},
in \textit{Quarks and leptons,
proceedings of the advanced study institute (Carg\`ese, Corsica, 1979)},
M.~L\'evy \textit{et al.}~eds.,
Plenum, New York, 1980;
\\
M.~Gell-Mann, P.~Ramond and R.~Slansky,
\textit{Complex spinors and unified theories},
in \textit{Supergravity},
D.Z.~Freedman and F.~van~Nieuwenhuizen eds.,
North Holland, Amsterdam, 1979;
\\
R.N.~Mohapatra and G.~Senjanovi\'c,
\textit{Neutrino mass and spontaneous parity violation},
\textit{Phys.~Rev.~Lett.} \textbf{44} (1980) 912.

\bibitem{leptogenesis1}
M.~Fukugita and T.~Yanagida,
\textit{Baryogenesis without grand unification},
\textit{Phys. Lett.} \textbf{B~174} (1986) 45.

\bibitem{leptogenesis2}
For reviews of leptogenesis see for instance
A.~Pilaftsis,
\textit{Heavy Majorana neutrinos and baryogenesis},
\textit{Int. J. Mod. Phys.} \textbf{A~14} (1999) 1811
[hep-ph/9812256]; \\
W.~Buchm\"uller and M.~Pl\"umacher,
\textit{Neutrino masses and the baryon asymmetry},
\textit{Int. J. Mod. Phys.} \textbf{A~15} (2000) 5047
[hep-ph/0007176]; \\
E.Ãƒ.~Paschos,
\textit{Leptogenesis},
\textit{Pramana} \textbf{62} (2004) 359
[hep-ph/0308261]; \\
S. Davidson, E. Nardi and Y. Nir,
\textit{Leptogenesis},
\textit{Phys. Rept.} \textbf{466} (2008) 105
[arXiv:0802.2962].

\bibitem{exp}
J.~Hosaka \textit{et al.},
Super-Kamiokande Collaboration, 
\textit{Three flavor neutrino oscillation analysis
of atmospheric neutrinos in Super-Kamiokande}, 
\textit{Phys. Rev.} \textbf{D~74} (2006) 032002
[arXiv:0604011];
\\
S.~Abe \textit{et al.},
KamLAND Collaboration,
\textit{Precision measurement of neutrino oscillation parameters
with KamLAND}, 
\textit{Phys. Rev. Lett.} \textbf{100} (2008) 221803
[arXiv:0801.4589];
\\
C.~Arpesella \textit{et al.},
Borexino Collaboration,
\textit{New results on solar neutrino fluxes
from 192 days of Borexino data},
\textit{Phys. Rev. Lett.} \textbf{101} (2008) 091302
[arXiv:0805.3843];
\\
B.~Aharmim \textit{et al.},
SNO Collaboration,
\textit{An independent measurement
of the total active $^8$B solar neutrino flux
using an array of $^3$He proportional counters
at the Sudbury Neutrino Observatory},
\textit{Phys. Rev. Lett.} \textbf{101} (2008) 111301
[arXiv:0806.0989];
\\
P.~Adamson \textit{et al.},
MINOS Collaboration,
\textit{Measurement of neutrino oscillations
with the MINOS detector in the NuMI beam},
\textit{Phys. Rev. Lett.} \textbf{101} (2008) 131802
[arXiv:0806.2237].

\bibitem{HPS}
P.F.~Harrison, D.H.~Perkins and W.G.~Scott,
\textit{Tri-bimaximal mixing and the neutrino oscillation data},
\textit{Phys. Lett.} \textbf{B~530} (2002) 167
[hep-ph/0202074].

\bibitem{TBM1}
See for instance
\\
E.~Ma,
\textit{$A_4$ symmetry and neutrinos with very different masses}, 
\textit{Phys. Rev.} \textbf{D~70} (2004) 031901
[hep-ph/0404199];
\\
G.~Altarelli and F.~Feruglio,
\textit{Tri-bimaximal neutrino mixing, $A_4$ and the modular symmetry}, 
\textit{Nucl. Phys.} \textbf{B~741} (2006) 215
[hep-ph/0512103];
\\
C.~Luhn, S.~Nasri and P.~Ramond,
\textit{Tri-bimaximal neutrino mixing
and the family symmetry $\mathcal{Z}_7 \rtimes \mathcal{Z}_3$},
\textit{Phys. Lett} \textbf{B~652} (2007) 27
[arXiv:0706.2341],
and references therein.

\bibitem{prescient}
W.~Grimus and L.~Lavoura,
\textit{A model realizing the Harrison--Perkins--Scott lepton mixing matrix},
\textit{J.~High Energy Phys.} \textbf{01} (2006) 018
[hep-ph/0509239].

\bibitem{tribi}
W. Grimus and L. Lavoura,
\textit{Tri-bimaximal mixing from symmetry only},
\textit{J. High Energy Phys.} \textbf{04} (2009) 013
[arXiv:0811.4766].

\bibitem{tri}
W.~Grimus and L.~Lavoura,
\textit{A model for trimaximal lepton mixing},
\textit{J. High Energy Phys.} \textbf{09} (2008) 106
[arXiv:0809.0226].

\bibitem{tri1}
W.~Grimus and L.~Lavoura,
\textit{A three-parameter neutrino mass matrix with maximal $CP$ violation},
\textit{Phys. Lett.} \textbf{B~671} (2009) 456
[arXiv:0810.4516].

\bibitem{manohar}
E.E. Jenkins and A.V. Manohar,
\textit{Tri-bimaximal mixing, leptogenesis, and $\theta_{13}$},
\textit{Phys. Lett.} \textbf{B~668} (2008) 210
[arXiv:0807.4176].

\bibitem{hagedorn}
E.~Bertuzzo, P.~Di Bari, F.~Feruglio and E.~Nardi, 
\textit{Flavor symmetries,
leptogenesis and the absolute neutrino mass scale},
\textit{J. High Energy Phys.} \textbf{11} (2009) 036
[arXiv:0908.0161];
\\
C.~Hagedorn, E.~Molinaro and S.T.~Petcov,
\textit{Majorana phases and leptogenesis
in seesaw models with $A_4$ symmetry},
\textit{J. High Energy Phys.} \textbf{09} (2009) 115
[arXiv:0908.0240];
\\
D.~Aristizabal Sierra, F.~Bazzocchi, I.~de~Medeiros Varzielas,
L.~Merlo and S.~Morisi,
\textit{Tri-bimaximal lepton mixing and leptogenesis},
arXiv:0908.0907.

\bibitem{nasri}
R.N.~Mohapatra, S.~Nasri and H.-B.~Yu,
\textit{$S_3$ symmetry and tri-bimaximal mixing},
\textit{Phys. Lett.} \textbf{B~639} (2006) 318
[hep-ph/0605020].

\bibitem{GL01}
L.~Lavoura and W.~Grimus,
\textit{Seesaw model with softly broken $L_e - L_\mu - L_\tau$},
\textit{J. High Energy Phys.} \textbf{09} (2000) 007
[hep-ph/0008020];
\\
W.~Grimus and L.~Lavoura,
\textit{Softly broken lepton numbers and maximal neutrino mixing},
\textit{J. High Energy Phys.} \textbf{07} (2001) 045
[hep-ph/0105212];
\\
W.~Grimus and L.~Lavoura, 
\textit{Softly broken lepton numbers:
an approach to maximal neutrino mixing},
\textit{Acta Phys. Pol.} \textbf{B~32} (2001) 3719 
[hep-ph/0110041].

\bibitem{albright}
C.A. Albright and W. Rodejohann,
\textit{Comparing trimaximal mixing and its variants with deviations
  from tri-bimaximal mixing},
\textit{Eur. Phys. J.} \textbf{C~62} (2009) 599
[arXiv:0812.0436].

\bibitem{fogli}
G.L.~Fogli, E.~Lisi, A.~Marrone and A.~Palazzo,
\textit{Global analysis of three-flavor neutrino masses and mixings},
\textit{Prog. Part. Nucl. Phys.} \textbf{57} (2006) 742
[hep-ph/0506083];
\\
G.L.~Fogli, E.~Lisi, A.~Marrone, A.~Palazzo and A.M.~Rotunno,
\textit{Hints of $\theta_{13} > 0$ from global neutrino data analysis},
\textit{Phys. Rev. Lett.} \textbf{101} (2008) 141801
[arXiv:0806.1649].

\bibitem{tortola}
T.~Schwetz, M.A.~T\'ortola, and J.W.F.~Valle,
\textit{Three-flavour neutrino oscillation update},
\textit{New J. Phys.} \textbf{10} (2008) 113011
[arXiv:0808.2016].

\bibitem{casas}
J.A.~Casas and A.~Ibarra, 
\textit{Oscillating neutrinos and $\mu \to e \gamma$}, 
\textit{Nucl. Phys.} \textbf{B~618} (2002) 171
[hep-ph/0103065].
\\
See also
G.C.~Branco, T.~Morozumi, B.M.~Nobre and M.N.~Rebelo, 
\textit{A bridge between CP violation at low energies and leptogenesis},
\textit{Nucl. Phys.}  \textbf{B~617} (2001) 475 
[hep-ph/0107164];
\\
M.N.~Rebelo, 
\textit{Leptogenesis without $CP$ violation at low energies}, 
\textit{Phys. Rev.} \textbf{D~67} (2003) 013008 
[hep-ph/0207236].

\bibitem{GL-lg}
W. Grimus and L. Lavoura,
\textit{Leptogenesis in seesaw models
with a two-fold degenerate neutrino Dirac mass matrix},
\textit{J. Phys.} \textbf{G~30} (2004) 1073
[hep-ph/0311362].

\end{thebibliography}
\end{document}